\documentclass[sigconf]{acmart}
\usepackage{multirow}

\newcommand\blfootnote[1]{%
  \begingroup
  \renewcommand\thefootnote{*}\footnotetext{#1}%
  \addtocounter{footnote}{-1}%
  \endgroup
}

\AtBeginDocument{%
  \providecommand\BibTeX{{%
    \normalfont B\kern-0.5em{\scshape i\kern-0.25em b}\kern-0.8em\TeX}}}

\copyrightyear{2020}
\acmYear{2020}
\setcopyright{acmcopyright}
\acmConference[SIGIR '20]{Proceedings of the 43rd International ACM SIGIR Conference on Research and Development in Information Retrieval}{July 25--30, 2020}{Virtual Event, China}
\acmBooktitle{Proceedings of the 43rd International ACM SIGIR Conference on Research and Development in Information Retrieval (SIGIR '20), July 25--30, 2020, Virtual Event, China}
\acmPrice{15.00}
\acmDOI{10.1145/3397271.3401332}
\acmISBN{978-1-4503-8016-4/20/07}

\begin{document}
\fancyhead{} 

\title{$L2R^2$: Leveraging Ranking for Abductive Reasoning}


\author{Yunchang Zhu, Liang Pang${}^{*}$, Yanyan Lan, Xueqi Cheng}
\affiliation{%
   \institution{CAS Key Lab of Network Data Science and Technology, \\ Institute of Computing Technology, Chinese Academy of Sciences, Beijing, China}
}
\affiliation{%
	\institution{University of Chinese Academy of Sciences, Beijing, China}
}
\email{{zhuyunchang17s, pangliang, lanyanyan, cxq}@ict.ac.cn}


\begin{abstract}\blfootnote{Corresponding author}
The abductive natural language inference task ($\alpha$NLI) is proposed to evaluate the abductive reasoning ability of a learning system.
In the $\alpha$NLI task, two observations are given and the most plausible hypothesis is asked to pick out from the candidates.
Existing methods simply formulate it as a classification problem, thus a cross-entropy log-loss objective is used during training.
However, discriminating true from false does not measure the plausibility of a hypothesis, for all the hypotheses have a chance to happen, only the probabilities are different.
To fill this gap, we switch to a ranking perspective that sorts the hypotheses in order of their plausibilities.
With this new perspective, a novel $L2R^2$ approach is proposed under the learning-to-rank framework.
Firstly, training samples are reorganized into a ranking form, where two observations and their hypotheses are treated as the query and a set of candidate documents respectively.
Then, an ESIM model or pre-trained language model, e.g. BERT or RoBERTa, is obtained as the scoring function.
Finally, the loss functions for the ranking task can be either pair-wise or list-wise for training.
The experimental results on the ART dataset reach the state-of-the-art on the public leaderboard.
\end{abstract}


\maketitle

\section{Introduction}
Abduction is considered to be the only logical operation that can introduce new ideas~\cite{peirce1965principles}.
It contrasts with other types of inference such as entailment, which refers to the well-known natural language inference tasks (NLI), that focuses on inferring only such information that is already provided in the premise.
Therefore, abduction reasoning is an important inference type deserved to be explored.
A new reasoning task, namely the abductive natural language inference task ($\alpha$NLI), is proposed to test the abductive reasoning capability of an AI system~\cite{Bhagavatula2020Abductive}.
Different from traditional NLI tasks, $\alpha$NLI first provides two pieces of narrative text treated as a start observation and an end observation. The most plausible explanation is then asked to pick out from the candidate hypotheses.

\begin{figure}
  \centering
  \includegraphics[width=\linewidth]{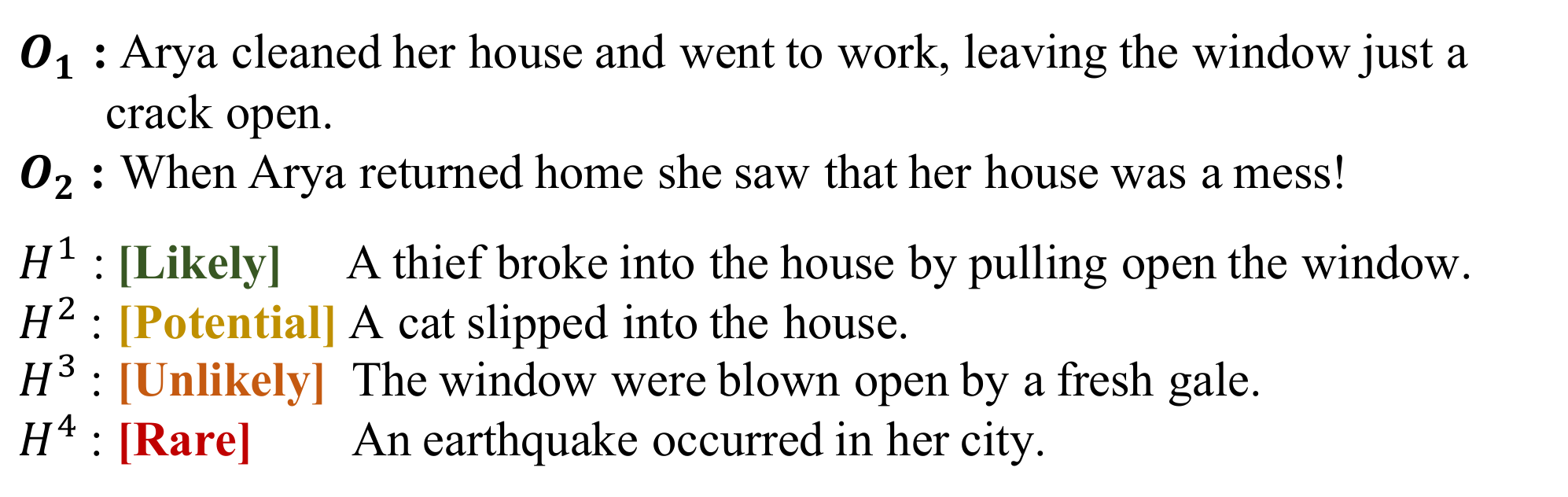}
  \caption{An example in the $\alpha$NLI task. Given observations, multiple hypotheses are plausible with their probabilities.}
  \label{fig:example}
\end{figure}

Many models have been successfully developed for the NLI tasks and directly adopted in the new proposed $\alpha$NLI task. 
These methods for NLI tasks treat the entailment between two sentences from a classification perspective, also treat $\alpha$NLI task as a binary-choice question answering problem, which selects one plausible hypothesis from two. 
However, discriminating true from false does not measure the plausibility of a hypothesis in abductive reasoning task, where all the hypotheses have a chance to happen with their probabilities, though some of their values are close to zero.
As we can see in Figure~\ref{fig:example}, from a tidy room (observation $O_1$) to a mess room (observation $O_2$), we do not know what has happened. Thus, four hypotheses are proposed, where `thief broke into the room' is the most likely happened, and `cat slipped into the room' is also a potential answer. Nevertheless, even for the hypothesis `earthquake' is also reasonable, just with a very small probability. It is hard to draw a line to determine which one is true from others.


Depending on these insights, we argue that $\alpha$NLI is better to be treated as a ranking problem. 
From the ranking perspective, binary-choice question answering setting in  the recent $\alpha$NLI task is just an incomplete pair-wise ranking scenario that only considers partial plausible order of given hypotheses.
In order to fully model the plausibility of the hypotheses, we switch to a complete ranking perspective and propose a novel learning to rank for reasoning ($L2R^2$) approach\footnote[1]{The source code is available at \url{https://github.com/zycdev/L2R2}.} for $\alpha$NLI task. 
$L2R^2$ adopts the mature learning-to-rank framework, which first reorganizes training instances into a ranking form. 
Specifically, two observations $O_1$ and $O_2$ can be view as a query, and the candidate hypotheses can be view as a set of candidate documents.
The relevance degree between a query and each document represents the plausible probability between observations and each hypothesis. 
Then, two parts of the learning-to-rank framework, scoring function and loss function are designed for $\alpha$NLI task.
Two types of scoring functions are chosen in this paper, the matching model ESIM~\cite{chen2017enhanced} and the pre-trained language models, e.g. BERT~\cite{devlin2019bert} and RoBERTa~\cite{liu2019roberta}. 
Besides, pair-wise and list-wise loss functions are applied to train the ranking task. 
The experimental results show that our $L2R^2$ approach achieves a new state-of-the-art accuracy on the blind test set of ART. 
Further analyses illustrate that the benefit of the ranking perspective is to assign a proper plausibility to each hypothesis, instead of either 0 or 1.

\section{Task Formalization}
The task of $\alpha$NLI contains two major concepts, observation and hypothesis. The observation describes the state of the scene, while the hypothesis is the imaginary cause that transforms one observation to another.
The famous Piaget's cognitive development theory tells us that our world is a dynamic system of continuous change, which involves transformations and states. 
Therefore, predicting the transformation is the core of the $\alpha$NLI task.

In detail, two observations are given $O_1, O_2 \in \mathcal{O}$, where $\mathcal{O}$ is the space of all possible observations. The goal of $\alpha$NLI task is to predict the most plausible hypothesis $H^* \in \mathcal{H}$, where $\mathcal{H}$ is the space of all hypotheses. Note that the happening time of observation $O_1$ is earlier than $O_2$.
Inspired by the traditional NLI task, where the hypothesis is regarded to be directly entailed from the premise.
However, the relation between hypothesis and two observations in $\alpha$NLI task is in a totally different way,
\begin{equation*}
  H^* = \arg\max_{H^j} P(H^j|O_1,O_2) = \arg\max_{H^j} P(O_2|O_1,H^j) P(H^j|O_1),
\end{equation*}
where hypothesis $H^j$ is depended on the first observation $O_1$, and the last observation $O_2$ is depended on $O_1$ and $H^j$. The best hypothesis $H^*$ is the one to max the score of these two parts.
It can be modeled by a scoring function that treats $O_1$, $O_2$ and $H^j$ as input, and outputs a real value $s_j$, e.g. scoring function: $f: \mathcal{O} \times \mathcal{H} \times \mathcal{O} \to \mathbb{R}$

For easy model adaptation, $\alpha$NLI in the ART dataset is originally defined as a binary-choice question answering problem, whose goal is to choose the most plausible hypothesis from two candidates $H^1$ and $H^2$.
From the classification perspective, it can be formalized as a discriminative task that distinguishes the category of $s_1-s_2$.
The positive indicates $H^1$ is more plausible than $H^2$, while the negative is the opposite.
We argue that it is an incomplete pair-wise approach in a ranking view, which only considers a small portion of the order in a ranking list and yields poor performance.

Therefore, we reformulate this task from the ranking perspective and adopt the learning-to-rank framework.
In this framework, observations $O_1$ and $O_2$ can be regarded as a query, and their candidate hypotheses $\mathbf{H}=\{H^j\}_{j=1}^{N}$ can be viewed as the corresponding candidate document set labeled with plausibility scores $\mathbf{y}=\{y_{j}\}_{j=1}^{N}$, where $N$ is the number of candidate hypotheses. 
The loss function is a key part of the learning-to-rank framework, where point-wise, pair-wise and list-wise are three commonly used loss function types.
In this paper, we only consider pair-wise and list-wise loss function, because point-wise is just a classification loss that does not take the order of the hypotheses into consideration.
Given the plausibility scores, we can make all possible hypotheses pairs, when plausibility scores are different, in order to train on a pair-wise loss function. 
We also use a list-wise loss function by treating the candidate hypotheses as an ordered list, which measures the error on a whole ranking list. 

\section{Our Approach}
Under the ranking formalization, we proposed our learning to rank for reasoning ($L2R^2$) approach, which is an implementation of the learning-to-rank framework for the $\alpha$NLI.
The learning-to-rank framework typically consists of two main components, e.g. a scoring function used to generate a real value score for a query-document pair and a loss function used to examine the accurate prediction of the ground truth rankings.

\subsection{Scoring Function}\label{scoring_func}
The scoring function $f$ can be implemented in different forms, for example, the deep text matching models and the pre-trained language models can be employed as the scoring functions.

ESIM is a strong NLI model that uses Bi-LSTM to encode tokens within each sentence and perform cross-attention on these encoded token representations, whose performance on entailment NLI is close to state-of-the-art. Thus, it is a good choice to implement as a scoring function. 
ESIM takes two sentences \textit{premise} and \textit{hypothesis} as input.
For $\alpha$NLI task, the concatenation of $O_1$ and $H$ is treated as the \textit{premise}, and $O_2$ is treated as the \textit{hypothesis}. ESIM outputs a scalar score indicating the relevance between them.

In scoring functions based on pre-trained language models such as BERT or RoBERTa.
For $\alpha$NLI task, the observations $O_1, O_2$ and hypothesis $H$ are first concatenated into a narrative story with a delimiter token and a sentinel token.
Then, it feeds into the pre-trained language model to get a contextual embedding for each token. After that a mean pooling is applied to obtain the feature vector of the observations-hypothesis pair $((O_1, O_2), H)$. Finally, a dense layer is stacked upon to get the plausible score $f(O_1,H,O_2)$.

\subsection{Loss Function and Inference}
Though the implementation can be different, all scoring functions are optimized by minimizing the empirical risk as follow:
\begin{equation}
  \mathcal{R}(f) = \frac{1}{m}\sum\nolimits_{i=1}^{m}L(\boldsymbol{y}^{(i)}, \boldsymbol{s}^{(i)}),
\end{equation}
where $L$ is the loss function utilized to evaluate the prediction scores $\boldsymbol{s^{(i)}}$ for a single query. 
Since point-wise loss functions are only suited for absolute judgment, we only explore pair-wise and list-wise loss functions in this work.

Pair-wise loss functions are defined on the basis of pairs of hypotheses whose labels are different, where ranking is reduced to a classification on hypotheses pairs. 
Here, the pairwise loss functions of Ranking SVM \cite{herbrich2000large}, RankNet \cite{burges2005learning} and LambdaRank \cite{burges2007learning} are used.

Hinge loss used in Ranking SVM and logistic (cross entropy) loss used in RankNet both have the following form:
\begin{equation} \label{eq:pairwise}
  L^{p}(\boldsymbol{y}, \boldsymbol{s}) = \sum\nolimits_{y_j > y_k} \phi(s_j - s_k),
\end{equation}
where the $\phi$ functions are hinge function ($\phi(z) = \max\{0,1-z\})$ and logistic function ($\phi(z) = \log(1 + e^{-z})$) respectively; $y_j>y_k$ means that $H^j$ ranks higher (is more plausible) than $H^k$ with regards to the query $(O_1, O_2)$.

Building upon RankNet, LambdaRank uses the logistic loss and adapts it by reweighing each hypotheses pair:
\begin{equation}\label{eq:pairwise_enhance}
  L^{p}(\boldsymbol{y}, \boldsymbol{s}) = \sum\nolimits_{y_j > y_k} \Delta \operatorname{NDCG}(j,k) \log(1 + e^{-(s_j - s_k)}),
\end{equation}
the $\Delta\operatorname{NDCG}(j,k)$ is the absolute difference between the NDCG values when the ranking positions of $H^j$ and $H^k$ are swapped,
\begin{displaymath}
  \Delta \operatorname{NDCG}(j,k) = |G(y_j)-G(y_k)|\cdot|D^{-1}(j)-D^{-1}(k)|,
\end{displaymath}
\begin{displaymath}
G(y)=(2^{y}-1)/{\operatorname{maxDCG}}, D(r)=\log(1+r),
\end{displaymath}
are gain and discount functions respectively, and maxDCG is a normalization factor per query.

Note that binary-choice classification baselines for $\alpha$NLI can be viewed as special cases of pairwise ranking methods when $N=2$.

Listwise loss functions are defined on the basis of lists of hypotheses. In this paper, the loss functions of ListNet \cite{cao2007learning}, ListMLE \cite{xia2008listwise} and ApproxNDCG \cite{qin2010general} are employed. 

In the ListNet approach, K-L divergence between the permutation probability distribution for the scoring function $P(\pi|\boldsymbol{s})$ and that for the ground truth $P(\pi|\boldsymbol{y})$ is used as the loss function, where $\pi \in \Pi$ denotes a permutation. Due to the huge size of $\Pi$, ListNet reduces the training complexity by using the marginal distribution of the first position and the K-L divergence loss then becomes
\begin{equation}\label{eq:listnet}
  L^l(\boldsymbol{y}, \boldsymbol{s}) = -\sum\nolimits_{j=1}^{N} \frac{e^{y_j}}{\sum_{k=1}^{N} e^{y_k}} \log\frac{e^{s_j}}{\sum_{k=1}^{N} e^{s_k}}.
\end{equation}

Different from ListNet, ListMLE uses the negative log likelihood of the ground truth permutation as the loss function,
\begin{equation}\label{eq:listmle}
  L^l(\boldsymbol{y}, \boldsymbol{s}) = -\log P(\pi_{\boldsymbol{y}}|\boldsymbol{s}),
\end{equation}
where $\pi_{\boldsymbol{y}}$ is he ground truth permutation.

ApproxNDCG optimizes approximate NDCG directly, and its loss function is then defined as follow:
\begin{equation}\label{eq:approx_ndcg}
\begin{aligned}
  L^l(\boldsymbol{y}, \boldsymbol{s}) = 1 - \sum\nolimits_{j=1}^{N} G(y_j) \big/ \log(1+\hat{\pi}(H^j)),\\
  \hat{\pi}(H^j) = 1 + \sum\nolimits_{u \neq j} e^{-(s_j-s_u)}\big/(1+e^{-(s_j-s_u)}).
\end{aligned}
\end{equation}
where $\hat{\pi}(H^j)$ is the approximation for $\pi(H^j)$ that indicates the position of $H^j$ in the ranking list $\pi$.

In inference stage, since original $\alpha$NLI task is to pick the more plausible one from two hypotheses, we can choose the hypothesis with highest score as the prediction result.

\begin{table*}
  \caption{Performances of $L2R^2$ and baselines on the development set of ART dataset.}
  \label{tab:dev_acc}
  \begin{small}
  \begin{tabular}{lccccccc}
    \toprule
    \multirow{2}{*}{Model}     & Binary-Choice   & \multicolumn{3}{c}{Pairwise Ranking} & \multicolumn{3}{c}{Listwise Ranking} \\
    \cmidrule(l){3-5} \cmidrule(l){6-8}
                               & Classification  & Logistic  & Hinge   & LambdaRank     & KLD      & Likelihood        & ApproxNDCG \\
    \midrule
    ESIM                        & 57.98          & 58.86     & 58.88    & 58.61         & 59.08    & \textbf{59.10}   & 58.81 \\
    BERT$_{\textrm{LARGE}}$     & 67.75          & 72.26     & 72.45    & 71.34         & 72.26    & \textbf{73.30}   & 71.80 \\
    RoBERTa$_{\textrm{LARGE}}$  & 85.64          & 87.86     & 87.92    & 87.79         & \textbf{88.45}    & 88.12   & 86.55 \\
    \bottomrule
  \end{tabular}
  \end{small}
\end{table*}

\begin{table}
  \caption{Performances of $L2R^2$ and baselines on the test set of ART dataset. All the results come from the leaderboard\protect\footnotemark[2].}
  \label{tab:test_acc}
  \begin{small}
  \begin{tabular}{lc}
    \toprule
    Model        & Accuracy  \\ 
    \midrule
    Random & 50.41 \\
    BERT$_{\textrm{BASE}}$ & 63.62 \\
    BERT$_{\textrm{LARGE}}$ & 66.75 \\
    RoBERTa$_{\textrm{LARGE}}$ & 83.91 \\
    McQueen + RoBERTa$_{\textrm{LARGE}}$ & 84.18 \\
    HighOrderGN + RoBERTa$_{\textrm{LARGE}}$ & 82.04 \\
    egel & 85.95 \\
    \midrule
    $L2R^2$ (RoBERTa$_{\textrm{LARGE}}$ + KLD) & \textbf{86.81} \\
    \midrule
    Human & 92.90 \\
    \bottomrule
  \end{tabular}
  \end{small}
\end{table}

\section{Experiments}
In this section, the experimental results on a public dataset are demonstrated to evaluate our proposed approaches.

\subsection{Experimental Settings}
We conduct our experiments on the ART~\cite{Bhagavatula2020Abductive} dataset.
ART is the first large-scale benchmark dataset for abductive reasoning in narrative texts. 
It consists of $\sim$20K pairs of observations with over 200K explanatory hypotheses, where observations are drawn from a collection of manually curated stories, and the hypotheses are collected by crowd-sourcing. 
Besides, the candidate hypotheses for each narrative context in the test sets are selected through an adversarial filtering algorithm that uses BERT$_{\textrm{LARGE}}$ as the adversary.

For our $L2R^2$ approach, the data need to reorganize into a ranking form.
Concretely, we merge original instances $(O_1,O_2,H^i,H^j)_{i \neq j}$ sharing the same observation pair $(O_1, O_2)$ into a new instance $(O_1,O_2,\mathbf{H})$, where $\mathbf{H}=\{H^{j}\}_{j=1}^{N}$ is a set of candidate hypotheses for a given observation pair. In the ART training set, there are an average of 13.41 hypotheses for each observation pair $(O_1, O_2)$, of which 4.05 are plausible.
We further employ a heuristic labeling strategy to construct ground truth plausibility scores $\mathbf{y}=\{y_{j}\}_{j=1}^{N}$ for $\mathbf{H}$. 
Consider $j$-th hypothesis $H^j$ for $(O_1,O_2)$, the ground truth plausibility score $y_j$ of $H^j$ is labeled with $\frac{\#(H^j\text{ occurs as plausible})}{\#(H^j\text{ occurs})}$.

To demonstrate the effectiveness of our approach, we develop 18 $L2R^2$ models based on three scoring functions, i.e. ESIM, BERT$_{LARGE}$ and RoBERTa$_{LARGE}$, with six ranking loss functions, including \textit{Logistic} for the loss (Eq~\ref{eq:pairwise}) used in RankNet, \textit{Hinge} for that (Eq~\ref{eq:pairwise}) used in Ranking SVM, \textit{LambdaRank} for that (Eq~\ref{eq:pairwise_enhance}) used in LambdaRank, \textit{KLD} for that (Eq~\ref{eq:listnet}) used in ListNet, \textit{Likelihood} for that (Eq~\ref{eq:listmle}) used in ListMLE, and \textit{ApproxNDCG} for that (Eq~\ref{eq:approx_ndcg}) used in ApproxNDCG.

Three binary-choice classification models are selected as our baselines. They have the same structures with the aforementioned scoring functions, whereas the only difference is that they are trained on original data with the cross entropy loss function.

For implementation details, we employ Adam as the optimizer and use early-stopping to select the best model. The models based on ESIM use 300 as the LSTM hidden size, which are trained for at most 64 epochs with batch size set to 32 and learning rate set to 4e-4. The models based on pre-trained language models are finetuned for at most 10 epochs with batch size set to 4 and learning rate set to 5e-6. 
The evaluation method \textit{accuracy} defined in \cite{Bhagavatula2020Abductive} is used.

\subsection{Experimental Results}
Table~\ref{tab:dev_acc} shows the experimental results of $L2R^2$ models and baselines on the development set. Our best model was evaluated officially on the test set, which achieved the state-of-the-art accuracy (Table~\ref{tab:test_acc}).

We summarize our observations as follows. 
(1) All 16 versions of our $L2R^2$ approaches improve the performance on the adbuctive reasoning task, which means that the ranking perspective is better than classification.
(2) Pair-wise models perform better than classification models, and most list-wise models perform better than pair-wise models. The former boost can be attributed to full version of pair-wise training, whereas the latter boost from pair-wise to list-wise is due to the global reasoning over the entire candidate set.
(3) BERT$_{LARGE}$ based ranking models have the largest gains about 8.2\% improvement over the corresponding baseline. It is because BERT$_{LARGE}$ was taken as the adversary for dataset construction, the substantial improvement illustrates that our $L2R^2$ approach is more robust to adversarial inputs.
(4) The loss functions optimizing NDCG metric, i.e. \textit{LambdaRank} and \textit{ApproxDNCG}, have poorer performances than others, mainly due to the gap between NDCG metric during training and accuracy metric during testing.

\subsection{Detailed Analyses}
\begin{figure}
  \centering
  \includegraphics[width=0.95\linewidth]{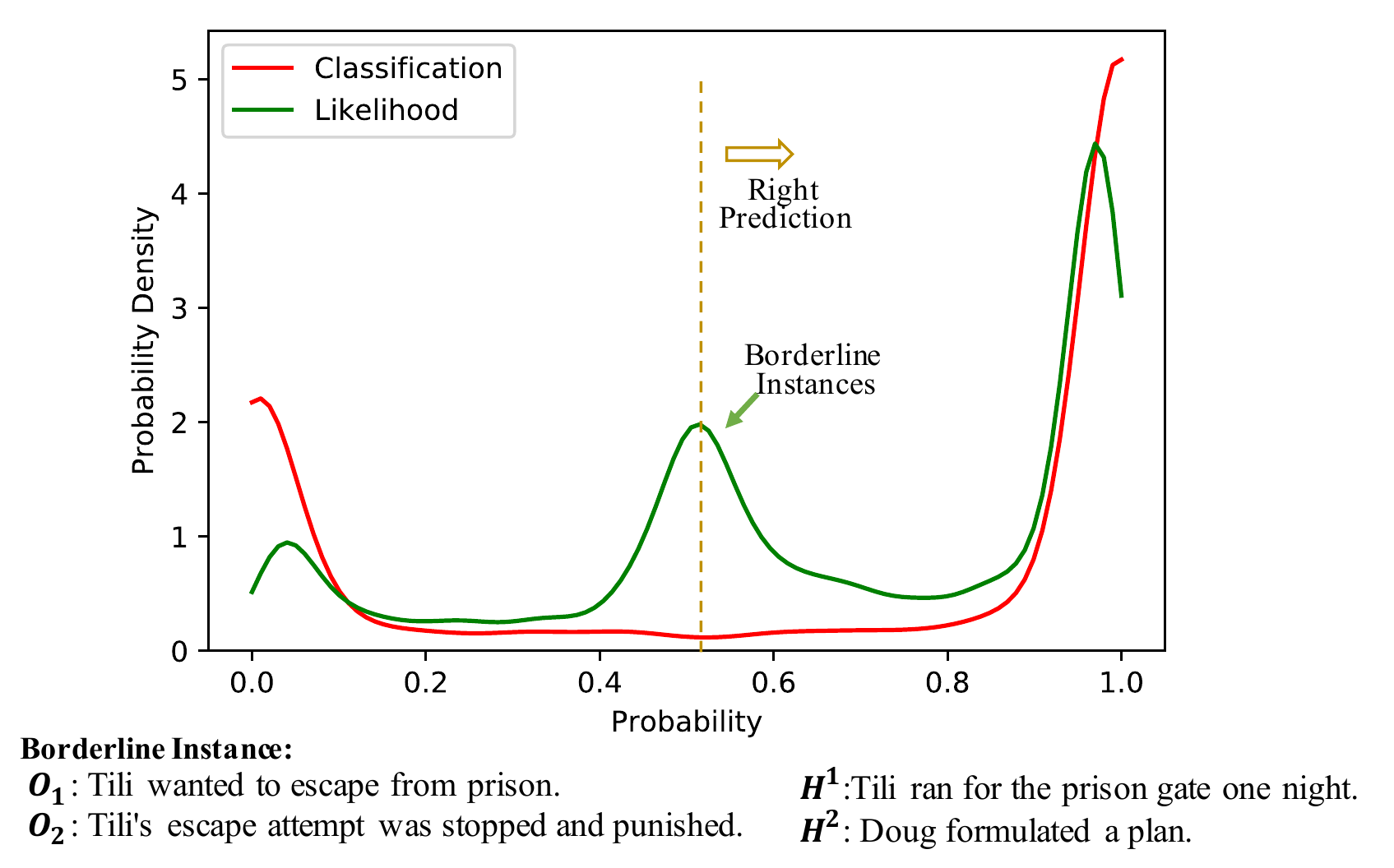}
\caption{Normalized probability distribution on the more plausible hypotheses in development set candidate pairs, predicted by an $L2R^2$ model and a classification model. Probabilities larger than 0.5 denote the right picked instances, while near 0.5 denote the borderline instances. }
  \label{fig:analysis}
\end{figure}

To further illustrate the rationality of our $L2R^2$ approach, Figure~\ref{fig:analysis} demonstrates two normalized score distributions on the more plausible hypotheses in development set candidate pairs, where the scores are predicted respectively by two models using BERT$_{LARGE}$ as the scoring function, the one trained with the classification loss and the other with list-wise likelihood loss.
The area under the curves in the right part (probability > 0.5) can be viewed as accuracy values. 
As shown in the figure, the classification model distinguishes the pairs of candidate hypotheses with a great disparity, either close to the probability 0 or 1, whereas the $L2R^2$ model has the ability to judge the borderline instances whose two candidates are competitive to each other.
Look at the sampled borderline instance in the bottom of Figure~\ref{fig:analysis}, where both hypotheses are likely to happen but $H^1$ is slightly more plausible, the $L2R^2$ model makes the right choice, which outputs two competitive probabilities for $H^1$ and $H^2$, 0.5891 vs 0.4109; whereas the classification model not only fails to distinguish which one is better but also outputs probabilities 0.0024 and 0.9976 in a significantly large gap.
That is to say, the ranking view in $L2R^2$ approach is a more reasonable way to model the abductive reasoning task.
\footnotetext[2]{See \url{https://leaderboard.allenai.org/anli/submissions/public}. Due to the limited submission times, only one of $L2R^2$ models was evaluated and achieved the best (2020-02-23).}

\section{Conclusion}
In the $\alpha$NLI task, all the hypotheses have their own chance to happen, so it is naturally treated as a ranking problem. 
From the ranking perspective, $L2R^2$ is proposed for the $\alpha$NLI task under the learning-to-rank framework, which contains a scoring function and a loss function. 
The experiments on the ART dataset show that reformulating the $\alpha$NLI task as ranking has improvements, also reaches the state-of-the-art performance on the public leaderboard.

\begin{acks}
This work was supported by Beijing Academy of Artificial Intelligence (BAAI) under Grants BAAI2020ZJ0303, the National Natural Science Foundation of China (NSFC) under Grants No. 61773362, and 61906180, the Youth Innovation Promotion Association CAS under Grants No. 2016102, the National Key R\&D Program of China under Grants No. 2016QY02D0405, the Tencent AI Lab Rhino-Bird Focused Research Program (No. JR202033).
\end{acks}

\bibliographystyle{ACM-Reference-Format}
\bibliography{main.bib}


\end{document}